\documentclass[10pt, a4paper, twocolumn]{article}

\usepackage[utf8]{inputenc}

\usepackage{color}
\usepackage{xcolor}
\usepackage{times}
\usepackage{graphicx}
\usepackage{multirow}
\usepackage{microtype}
\usepackage{comment}
\usepackage{subcaption}
\usepackage[none]{hyphenat} 
\usepackage{url}
\newcommand{\red}{\textcolor{red}}
\usepackage{sectsty}
\usepackage{cite}

\sectionfont{\fontsize{14}{15}\selectfont}
\subsectionfont{\fontsize{12}{15}\selectfont}

\usepackage[
top=0cm,
bottom=1.78cm,
left=1.65cm,
right=1.65cm,
headheight=2.8cm,
headsep=0cm,
includehead
]{geometry}

\usepackage{authblk}

\renewenvironment{quote}{%
   \list{}{%
     \leftmargin0.1cm   
     \rightmargin\leftmargin
   }
   \item\relax
}
{\endlist}

\newenvironment{sciabstract}{%
\begin{quote} \bf}
{\end{quote}}

\title{Detection of COVID-19 Using Heart Rate and \\ Blood Pressure: Lessons Learned from Patients with ARDS}

\author[1]{Milad Asgari Mehrabadi}
\author[1]{Seyed Amir Hossein Aqajari}
\author[2]{Iman Azimi}
\author[3]{Charles A Downs}
\author[1,4]{Nikil Dutt}
\author[1,4,5]{Amir M Rahmani}
\affil[1]{Department of Electrical Engineering and Computer Science, University of California Irvine, USA}
\affil[2]{Department of Computing, University of Turku, Finland}
\affil[3]{School of Nursing and Health Studies, University of Miami, USA}
\affil[4]{Department of Computer Science, University of California Irvine, USA}
\affil[5]{School of Nursing, University of California Irvine, USA}

\date{}

\begin{document} 




\maketitle 


\begin{sciabstract}
The world has been affected by COVID-19 coronavirus. At the time of this study, the number of infected people in the United States is the highest globally (7.9 million infections). Within the infected population, patients diagnosed with acute respiratory distress syndrome (ARDS) are in more life-threatening circumstances, resulting in severe respiratory system failure. 
Various studies have investigated the infections to COVID-19 and ARDS by monitoring laboratory metrics and symptoms. Unfortunately, these methods are merely limited to clinical settings, and symptom-based methods are shown to be ineffective. In contrast, vital signs (e.g., heart rate) have been utilized to early-detect different respiratory diseases in ubiquitous health monitoring. We posit that such biomarkers are informative in identifying ARDS patients infected with COVID-19. 
In this study, we investigate the behavior of COVID-19 on ARDS patients by utilizing simple vital signs. We analyze the long-term daily logs of blood pressure and heart rate associated with 70 ARDS patients admitted to five University of California academic health centers (containing 42506 samples for each vital sign) to distinguish subjects with COVID-19 positive and negative test results. In addition to the statistical analysis, we develop a deep neural network model to extract features from the longitudinal data. Using only the first eight days of the data, our deep learning model is able to achieve 78.79\% accuracy to classify the vital signs of ARDS patients infected with COVID-19 versus other ARDS diagnosed patients.

\end{sciabstract}
Keywords: COVID-19; ARDS; Wearable; Machine Learning; Internet-of-Things

\section{Introduction}
The acute respiratory distress syndrome (ARDS) is a life-threatening consequence of infection with SARS-CoV-2, the novel coronavirus that causes COVID-19 \cite{li2020acute}. ARDS is characterized by an overwhelming immune response and non-cardiogenic pulmonary edema that compromise gas exchange, resulting in severe respiratory failure. ARDS mortality ranges from 40\%-60\%; however, it is unclear if it is substantially higher if associated with COVID-19 infection, as it varies from 28.8\%-62\% 
\cite{li2020acute,tang2020comparison}. Currently, more than 38 million people worldwide have been infected with SARS-CoV-2 \cite{jhcovid}. In the United States, 7.9 million people have been infected with over 216,000 deaths \cite{jhcovid}. The impact of the COVID-19 pandemic is considerable and efforts to mitigate its spread through early detection cannot be over-emphasized.

Infections to COVID-19 have been conventionally investigated in clinical settings by monitoring laboratory metrics and symptoms \cite{jehi2020individualizing,callahan2020estimating}. These studies have focused on a large amount of subjective questionnaires and invasive laboratory test results. 
For example, Jehi \textit{et al.} \cite{jehi2020individualizing}  used
a large number of features  extracted from demographics, comorbidities, immunization history, symptoms, travel history, laboratory vairables, and medications to predict the infection with COVID-19. 
Li \textit{et al.} \cite{li2020acute} show that the oxygenation index and respiratory system compliance could be leveraged to study ARDS patients infected with COVID-19. Force \textit{et al.} \cite{force2012acute} propose that ARDS caused by factors rather than COVID-19 results in reduced lung compliance. However, reduced lung compliance in ARDS is typical of the disease \cite{li2020acute}. 

Such diagnostics are the gold standard methods to investigate COVID-19 and ARDS patients; however, they are limited to hospitals and clinical settings. Moreover, subjective symptom-based analyses were shown to be an ineffective strategy to qualify an individual's likelihood of contracting COVID-19 \cite{callahan2020estimating}. In contrast, various studies showed that vital signs such as heart rate and blood pressure could be exploited for early detection of infections and respiratory diseases \cite{shashikumar2017early}. 
We posit that such biomarkers are informative in identifying ARDS patients infected with COVID-19.
These biomarkers can be collected continuously and remotely due to the recent advancements in wearable electronics and Internet-of-Things-based devices. Therefore, the effectiveness of these biomarkers in early COVID-19 detection extends the monitoring services to remote settings. Understanding and leveraging these clinical measurements also play a significant role in preventive care and treatments \cite{matsumura2020comparison,gattinoni2020covid}.


Recognition of COVID-19 infections using big sensory data necessitates novel modeling and analysis techniques. The state-of-the-art studies often use traditional statistical models to predict COVID-19 infections. These studies have mostly studied the linear statistical relationship and association between the health parameters or extracted features from the subject's demographics, symptoms, laboratory tests, and medications \cite{jehi2020individualizing,callahan2020estimating}. For example, a full multi-variable logistic model is constructed in \cite{jehi2020individualizing} to predict COVID-19 using extracted features. However, such data with complex intensive longitudinal structure and temporal characteristics need to be investigated using nonlinear and advanced methods.  Machine learning algorithms, including Artificial Neural Networks, can be tailored in this regard to extract linear/nonlinear correlations in the data throughout the health monitoring.  


In this paper, we investigate the behavior of COVID-19 on ARDS patients by utilizing three longitudinal features: systolic and diastolic blood pressure and heart rate. We compare individuals who developed ARDS with and without COVID-19 to assess potential markers that could be used in early detection and prevention strategies. 
We use the University of California COVID Research Data Set (UC-CORDS) \cite{UCCORD} that contains comprehensive, structured information from patients admitted to the University of California Health's five academic health centers (i.e., UC Davis Health, UC San Diego Health, UC Irvine Health, UCLA Health, and UCSF Health). Moreover, we utilize statistical features and neural networks to distinguish between ARDS caused by COVID-19 and other factors. 
The biomarkers investigated in this study (i.e., heart rate and blood pressure) have the potential for scalable COVID-19 prevention and monitoring in everyday settings, thanks to the
ubiquitous availability of inexpensive, non-invasive portable and wearable sensors.
For instance, the Omron\textsuperscript{\tiny\textregistered} HeartGuide wrist-band \cite{omron} is an FDA-cleared example of such wearable devices capable of monitoring all these three biomarkers.
These observations have potential applications across community settings and for those living in collective housing.

\section{Results}\label{sec:results}
In this section, we discuss the results obtained by statistical analysis and neural networks.
\subsection{Statistical Observations}
We measured basic statistical features over BP and HR and compared them with COVID-19 test results. Table \ref{tbl:stat} shows the Point Biserial correlation between these features and age with the test results, and Table \ref{tbl:stat_CI} represents 95\% confidence interval (CI) of these features for each test group.

\begin{table}[!h]
\centering
\caption{Point Biserial correlation of statistical features and the test results (* shows significant correlation).}
\label{tbl:stat}
\resizebox{0.6\columnwidth}{!}{%
\begin{tabular}{|c|c|c|c|}
\hline
 \multicolumn{2}{|c|}{} & Correlation & {\it P}-value \\
\hline
\multirow{4}{*}{HR} & mean & -0.21 & .07 \\
                    & std & 0.14 & .24\\
                    & min & -0.25 & .031* \\
                    & max & -0.10 & .36 \\
\hline
\multirow{4}{*}{DBP} & mean & 0.11 & .33 \\
                     & std & -0.08 & .47\\
                     & min & 0.22 & .065 \\
                     & max & -0.26 & .027* \\
\hline
\multirow{4}{*}{SBP} & mean & 0.01 & .89 \\
                     & std & 0.04 & .70\\
                     & min & 0.10 & .36 \\
                     & max & -0.06 & .61 \\
\hline
Age & & -0.004 & 0.96 \\
\hline
\end{tabular}
}
\end{table}

\begin{table}[!h]
	\centering
	\caption{95\% confidence interval of each statistical feature for test group.}
	\label{tbl:stat_CI}
	\resizebox{0.8\columnwidth}{!}{%
		\begin{tabular}{|c|c|c|c|}
			\hline
			\multicolumn{2}{|c|}{} & 95\% CI (Positive) & 95\% CI (Negative) \\
			\hline
			\multirow{4}{*}{HR} & mean & 75.78-86.18 & 82.60-91.36 \\
			& std & 13.50-17.39 & 11.71-15.85\\
			& min & 44.04-52.07 & 50.19-60.28 \\
			& max & 123.17-144.57 & 129.36-152.74 \\
			\hline
			\multirow{4}{*}{DBP} & mean & 55.09-60.47 & 52.68-58.90 \\
			& std & 7.83-9.56 & 8.07-10.36\\
			& min & 29.56-36.93 & 23.96-23.19 \\
			& max & 91.83-110.78 & 107.60-150.02 \\
			\hline
			\multirow{4}{*}{SBP} & mean & 113.95-120.83 & 111.04-122.78 \\
			& std & 17.56-21.47 & 17.07-20.94\\
			& min & 53.59-69.59 & 49.17-64.19 \\
			& max & 180.64-204.98 & 180.86-215.91 \\
			\hline
			Age & & 55.50-66.62 & 55.94-66.48 \\
			\hline
		\end{tabular}
	}
\end{table}

\begin{figure}[!h]
\centering
\includegraphics[width=0.45\textwidth]{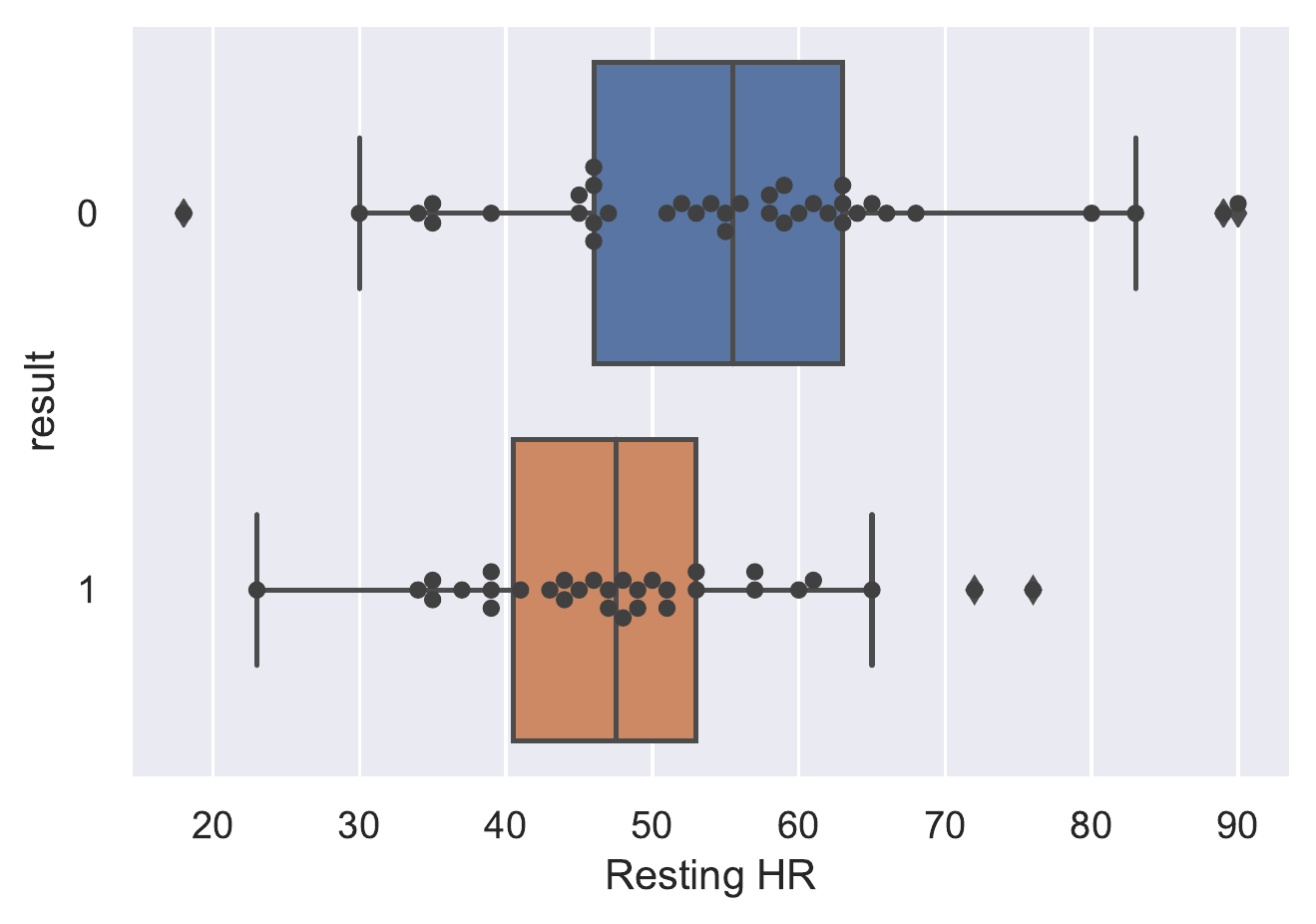}
\caption{Box plot of resting HR per each test result group.}
\label{fig:min_hr}
\end{figure}

Table \ref{tbl:stat} suggests significant negative correlations between the resting HR (min HR), max value of DBP, and test results. Fig. \ref{fig:min_hr} illustrates the difference in the distribution of resting HR between the positive and negative test result groups. The average resting HRs were 55.23 and 48.06 for the negative and positive test groups, respectively. Although resting HR shows a significant correlation with the test results, there is an overlap in the distribution of such a feature between positive and negative results. 


\subsection{Neural Network}
Due to the longitudinal aspect of the data, we consider a deep neural network architecture to predict the test results by only looking at BP and HR.  
The accuracy of this model reached as high as 74.32\% for the entire test data. 

\begin{figure*}[!t]
\centering
\begin{subfigure}{.45\textwidth}
  \centering
  \includegraphics[width=\linewidth]{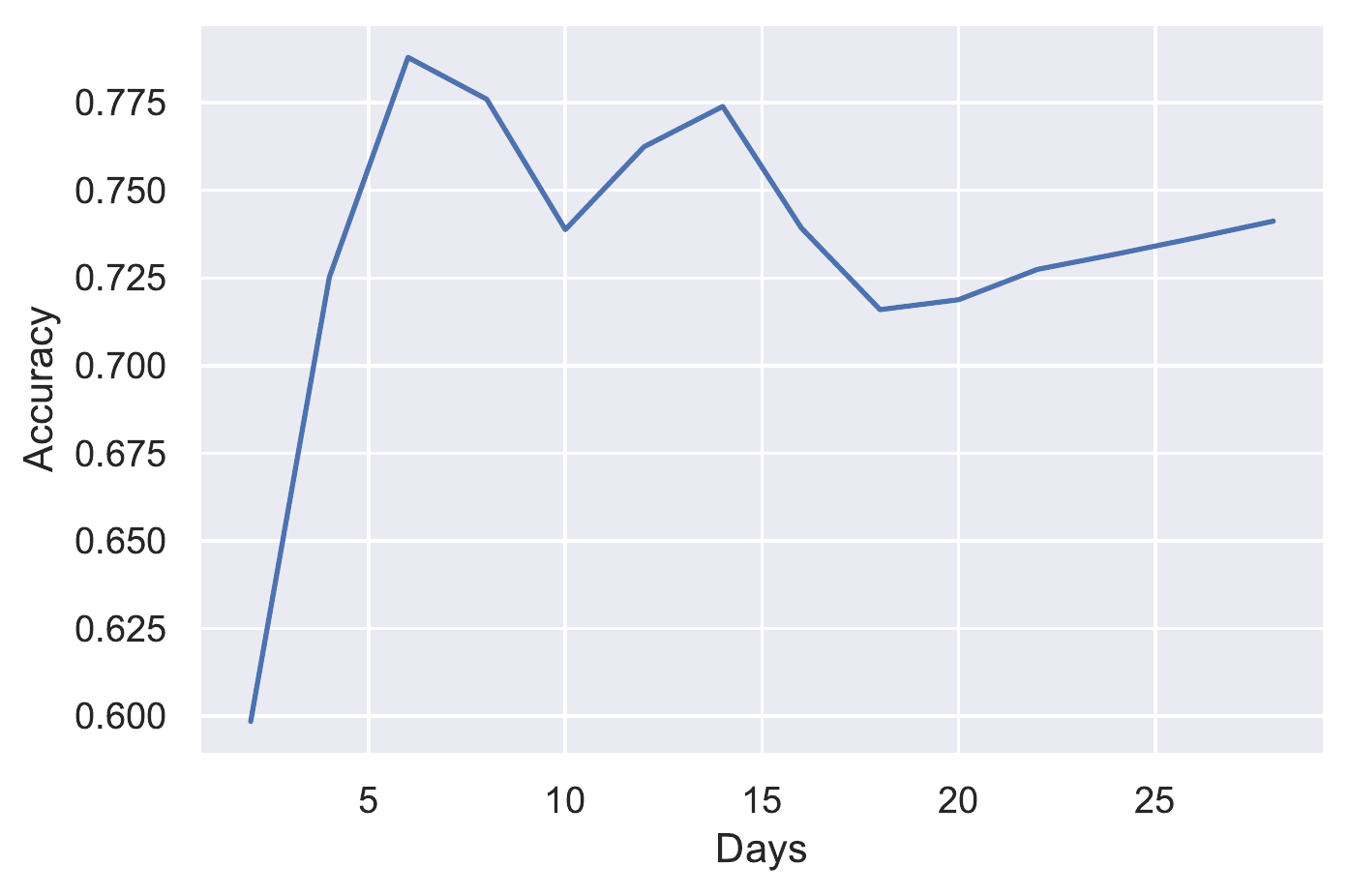}  
  \caption{}
  \label{fig:perf-first}
\end{subfigure}
\begin{subfigure}{.45\textwidth}
  \centering
  \includegraphics[width=\linewidth]{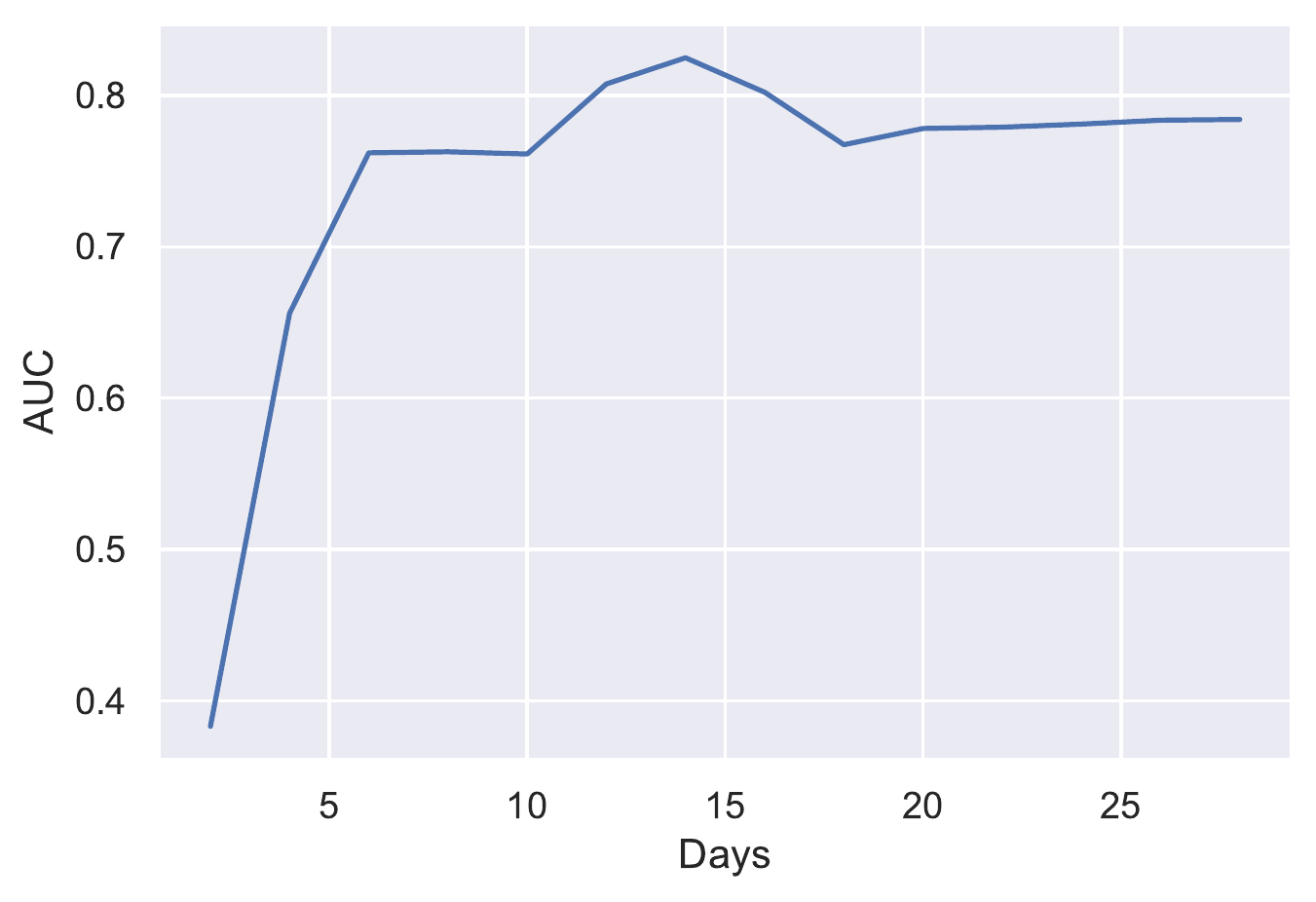}  
  \caption{}
  \label{fig:perf-second}
\end{subfigure}
\caption{The performance of the model in terms of accuracy (a) and AUC (c) using test data with respect to the number of included days.}
\label{fig:perf}
\end{figure*}

Besides, we tested the model with different testing sizes, which is extracted including N days ($N\in\{2, 4, ..., 28\}$), to see the model's performance by looking only at a limited number of days. Fig. \ref{fig:perf-first} shows the accuracy of the classification model with respect to days. Fig. \ref{fig:perf-second} illustrates the corresponding area under the curve (AUC) with given days. Fig. \ref{fig:perf-first} shows an increase in the model's accuracy at the beginning, starting from 59.85\% and reaching as high as 78.79\% on $8^{th}$ day. 

\begin{figure*}[!t]
\centering
\begin{subfigure}{.45\textwidth}
  \centering
  \includegraphics[width=\linewidth]{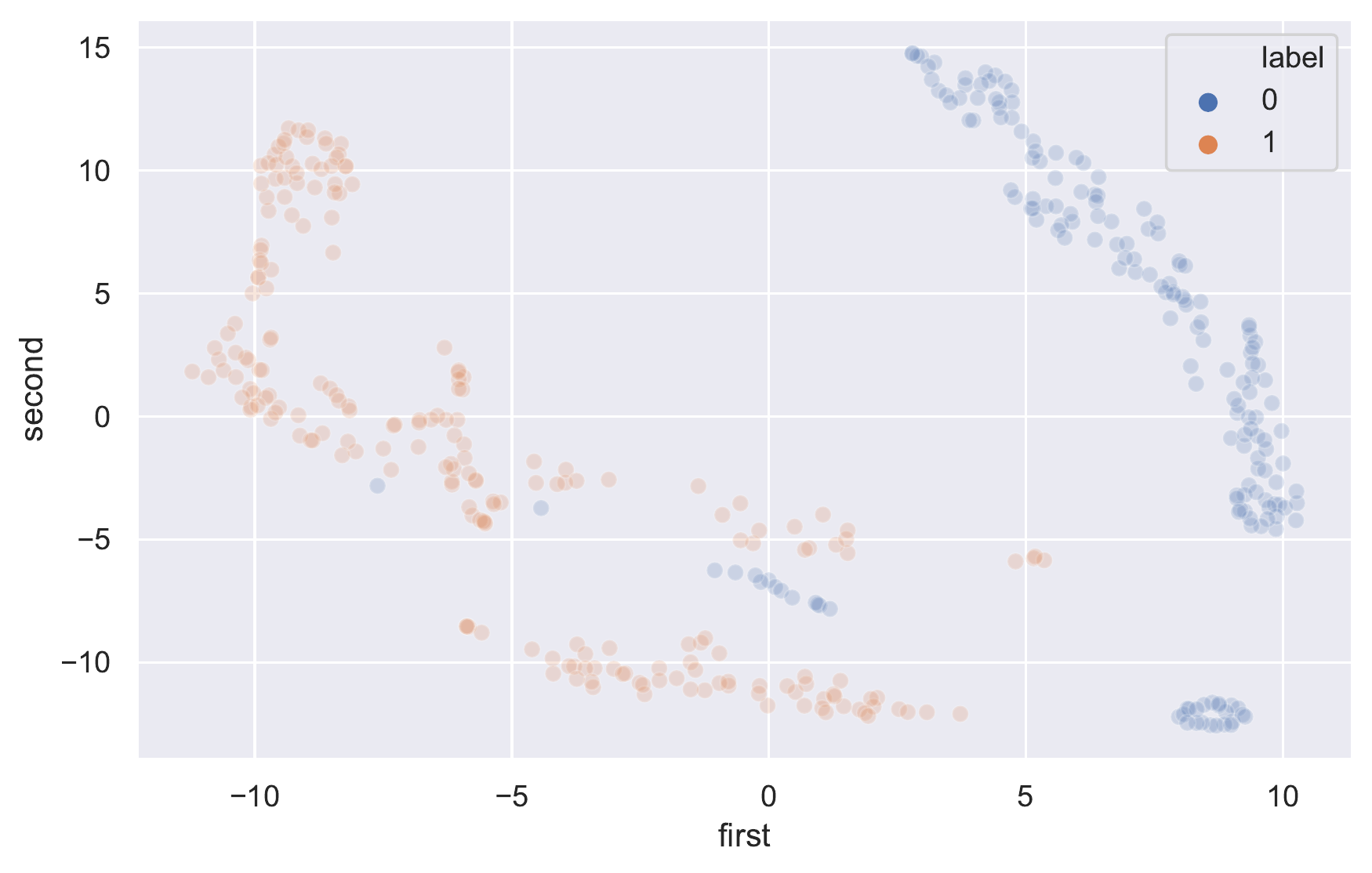}  
  \caption{}
  \label{fig:tsne-first}
\end{subfigure}
\begin{subfigure}{.45\textwidth}
  \centering
  \includegraphics[width=\linewidth]{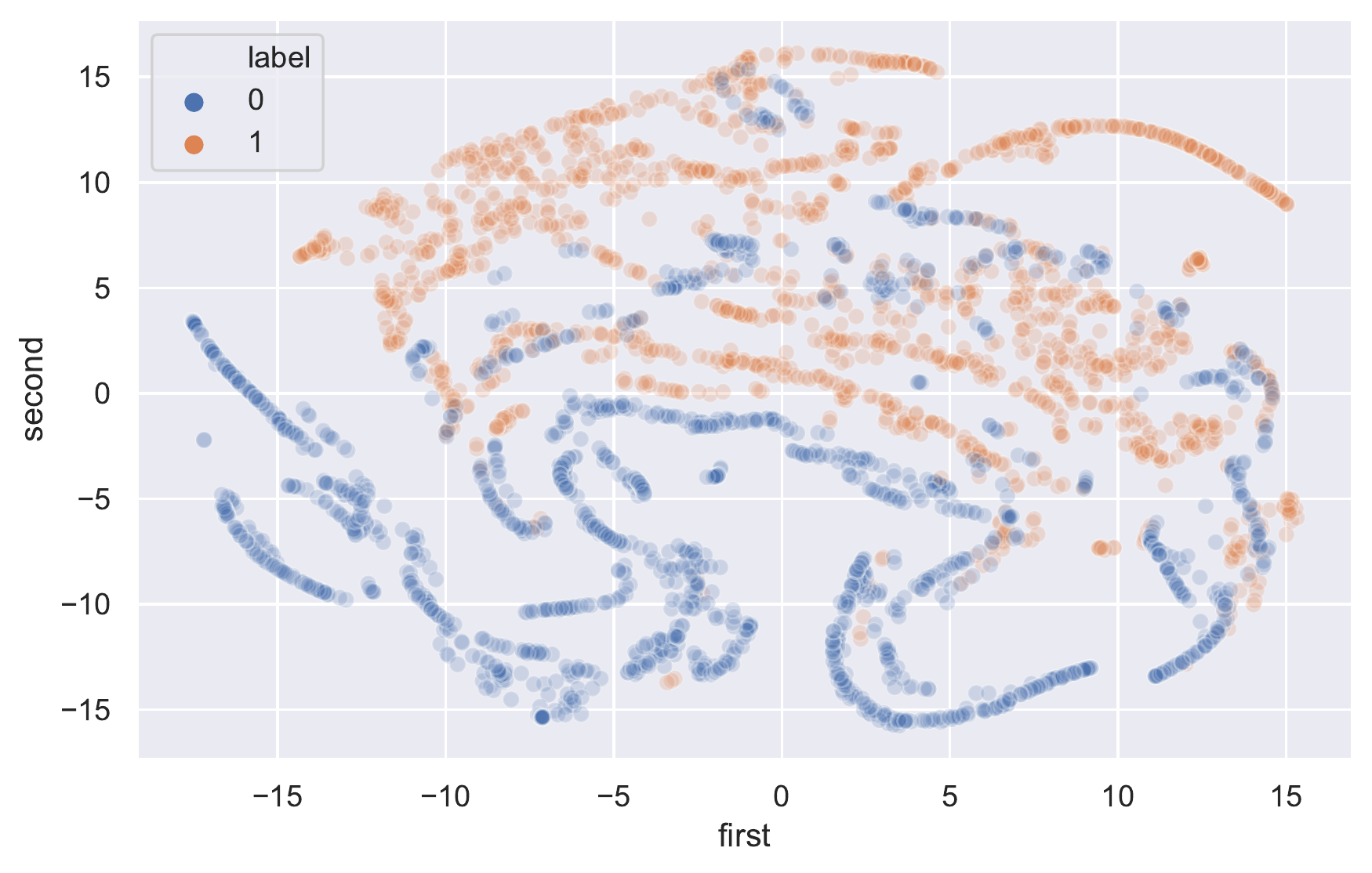}  
  \caption{}
  \label{fig:tsne-second}
\end{subfigure}
\caption{The 2-dimensional representation of test data using t-SNE considering 2 (a), and the entire test data (c). The colors represent the test label.}
\label{fig:tsne}
\end{figure*}

To better visualize the extracted features using neural networks, we used the t-SNE method \cite{maaten2008visualizing} to reduce the feature space dimension to two. We performed this method on the output of the dense layer with 100 neurons. Fig. \ref{fig:tsne} visualizes the test data with different included days. Fig. \ref{fig:tsne-first} shows that using extracted features by the deep neural network, the positive and negative cases are almost separated.  As the number of samples increases, the decision boundary calculation would be more challenging; however, the clusters are still distinguishable (Fig. \ref{fig:tsne-second}).

\section{Discussion}\label{sec:discussion}
A few of our observations warrant additional discussion. 
First, monitoring of blood pressure and heart rate may provide a useful strategy for individuals living in collective communities, such as nursing homes or rehabilitation facilities, as well as for healthy community-dwelling adults.  The potential impact could be to mitigate the spread of COVID-19, as well as allowing early detection of complications associated with infection, such as those at greater risk for ARDS. 

Second, we assessed for the presence of comorbidities in COVID-19 positive patient with ARDS, and reported that comorbid diagnoses such as type 2 Diabetes Mellitus, hyperglycemia, chronic obstructive pulmonary disease, elevated transaminase, and lactic acid dehydrogenase, bradycardia, acute ST segment elevation myocardial infarction, and metabolic derangements were more prevalent (data not shown).  This observation is in-line with other reports \cite{li2020acute, tang2020comparison, wu2020risk} demonstrating increased vulnerability among those with chronic health conditions, as well as reported metabolic derangements observed with COVID infection, especially among adults over 60 years of age. 

Third, there are other potential applications in modeling COVID-19. Specifically, there has been a discussion of how early COVID-19 arrived in the United States; the first cases were reported in California. It would be possible to review data prior to the first reported cases in the U.S. to validate the presence or absence of COVID-19 in our communities prior to January 2020. This is of importance as the viral genome sequence was confirmed in late January 2020, which allowed for the use of polymerase chain reaction to detect viral genetic material \cite{pasteur}. Antibody testing, which has been shown to be inconsistent, was used in the preceding months, raising the question of how early was COVID-19 in the United States.

There are related studies in the literature that propose prediction models for the patient's infection with COVID-19 in lab setups. Jehi \textit{et al.} \cite{jehi2020individualizing} created a statistical model to accurately predict infection with COVID-19 using the data from 11672 patients, tested before April 2, 2020. A full multi-variable logistic model was initially constructed to predict COVID-19 using features extracted from demographics, comorbidities, immunization history, symptoms, travel history, laboratory variables, and medications before testing. Although their c-index ranged from 0.839 to 0.863, their statistical model requires a broad set of features to predict a patient's infection with COVID-19. One of the drawbacks in their work is that some of these features can only be measured in clinical laboratory settings. 

In contrast, we considered two easily accessible features as well as utilizing a deep learning method to capture the short- and long-term dependencies in the time series data. There is a correlation between the simple statistical features and the test results. However, simple logistic regression models are insufficient due to the overlap in the feature space. Leveraging the nonlinear features extracted from our proposed neural network, we distinguished negative and positive COVID-19 test results with the AUC as high as 0.83 by using only blood pressure and heart rate values.

Moreover,  Callahan \textit{et al.} \cite{callahan2020estimating} investigated whether symptom-based screening is feasible in prioritized testing. To access feasibility, they started with predicting participants' test results with diagnoses of common respiratory viruses to co-infect patients positive for SARS-Cov-2 at Stanford Healthcare. They evaluated symptoms mentioned in clinical notes at the time the test was performed. For the respiratory viruses, AUC for the receiver operator curve on the test data ranged from 0.60 to 0.77. Based on their studies, they concluded that two of non-SARS-Cov2 viruses (i.e., influenza type A and RSV) were moderately predictable given presenting symptoms. However, SARS-Cov-2 and remaining common respiratory viruses were not highly predictable (AUROCs below 0.70). According to the model, although they suggested that symptom-based screening is an ineffective strategy to predict person's infection with COVID-19, the usage of vital signs (i.e., heart rate and blood pressure) is not investigated.

On the contrary, in this study, we mainly focused on ARDS patients as a population. Although our findings are only based on this population, these achievements could potentially lead future directions of our research to investigate the aforementioned vital signs for COVID-19 prediction tasks with other populations as well. Besides, another category of detection models focuses on identifying the characteristics of the patients with COVID-19 at a specific point in time, which usually is 1-2 months \cite{bhatraju2020covid,goyal2020clinical,richardson2020and}. Some of these studies aim to use the identified characteristics to predict critical cases of COVID-19; specifically, those most likely require hospitalization or Intensive care unit (ICU) admission. In \cite{carlino2020predictors, grasselli2020baseline, wang2020clinical}, the authors attempted to find the best predictors of ICU admission among infected patients with COVID-19.

In conclusion, we investigated the non-linear patterns in simple vital signs, namely, blood pressure and heart rate, which can be easily and reliably measured without the need for skilled medical professionals, in ARDS patients with positive and negative COVID-19 test results. Our proposed neural network-based model achieved 78.79\% accuracy, considering only the first eight days of data.
Using such prediction methods, the number of visits to the hospitals or care sites, as well as the chance of virus spread, can be reduced. Using wearable devices, it is possible to monitor vital signs of subjects in everyday settings without visiting a hospital or a care site. Utilizing the proposed model allows early detection of COVID-19 cases in free-living conditions. 

\section{Methods}\label{sec:objective}

\subsection{Data Set}
Data set plays an essential role in any prediction tasks. UC-CORDS data set provides comprehensive, structured information of patients admitted to the hospital at the University of California Health's five academic health centers (i.e., UC Davis Health, UC San Diego Health, UC Irvine Health, UCLA Health, and UCSF Health). This data set provides a wide range of information, including different observations, measurements and COVID-19 test results of patients. 

Notably, the vital signs are recorded every 15, 30, or 60 minutes based on the time during a day. For this study, we aimed to select participants with ARDS hospitalized after January $1^{st}$, 2020. The earliest available data in UC-CORDS for positive COVID-19 inpatients with ARDS was March $21^{st}$, 2020. Therefore, we only considered hospitalized ARDS (IDs 4195694 and 4191650 from SNOMED vocabulary \cite{SNOMED}) patients tested between March $21^{st}$, 2020 and August $1^{st}$, 2020. Since the number of observations with negative COVID-19 test results was more than positives, we considered patients with negative test results after July $1^{st}$, 2020. This re-sampling resulted in a more balanced data set (i.e., 19449 data points for each feature in the positive group and 23057 samples in the negative group). 
As of August $1^{st}$ 2020, this led to 32 and 38 participants with positive and negative test results, respectively. Table \ref{tbl:age} shows the age distribution of patients per each test result.

\begin{figure}[!h]
\centering
\includegraphics[width=0.45\textwidth]{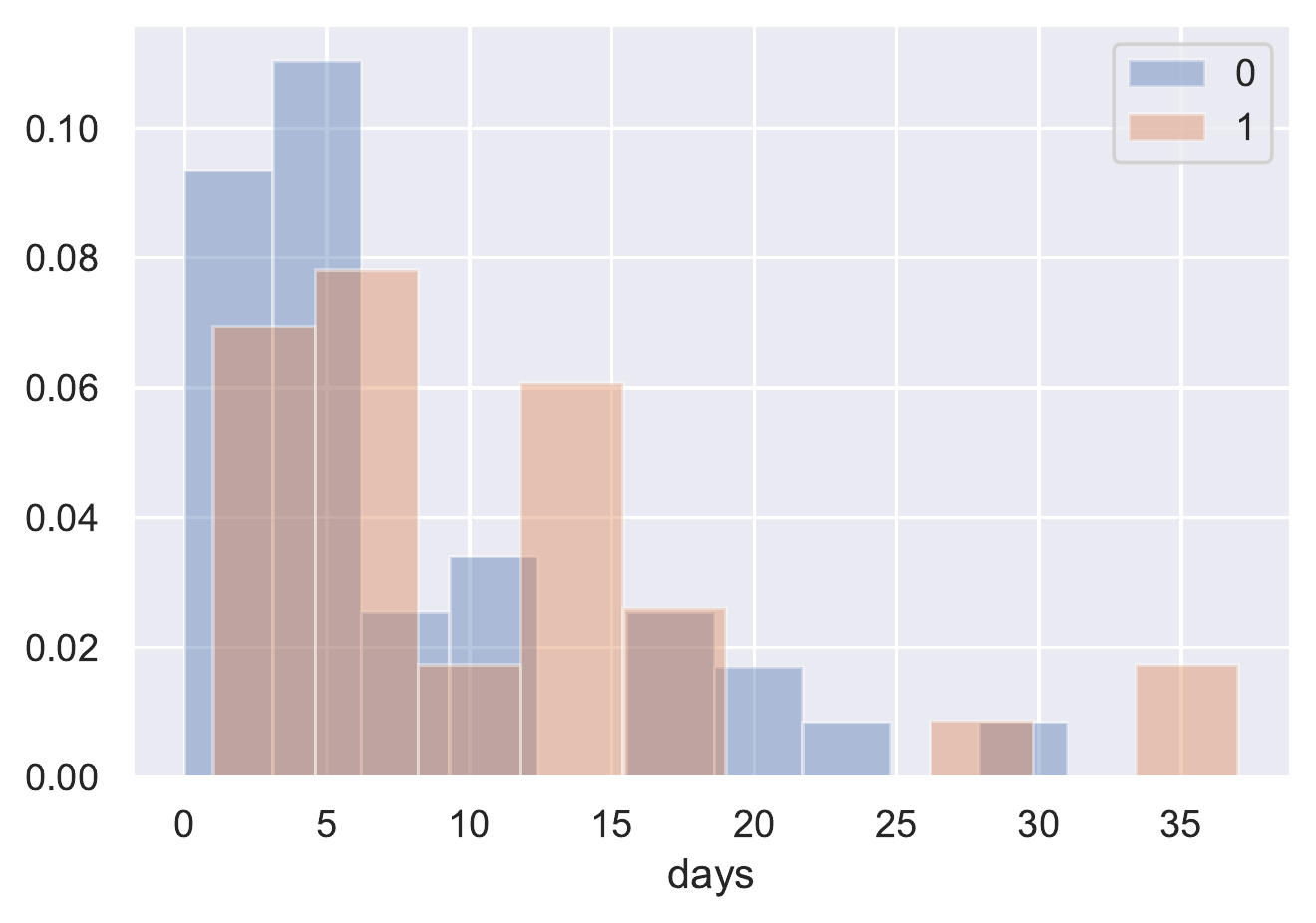}
\caption{Duration of available data in days for each population group. The value of 0 represents patients with negative test and 1 shows participants with positive test results.}
\label{fig:duration}
\end{figure}

\begin{figure}[!h]
\centering
\includegraphics[width=0.45\textwidth]{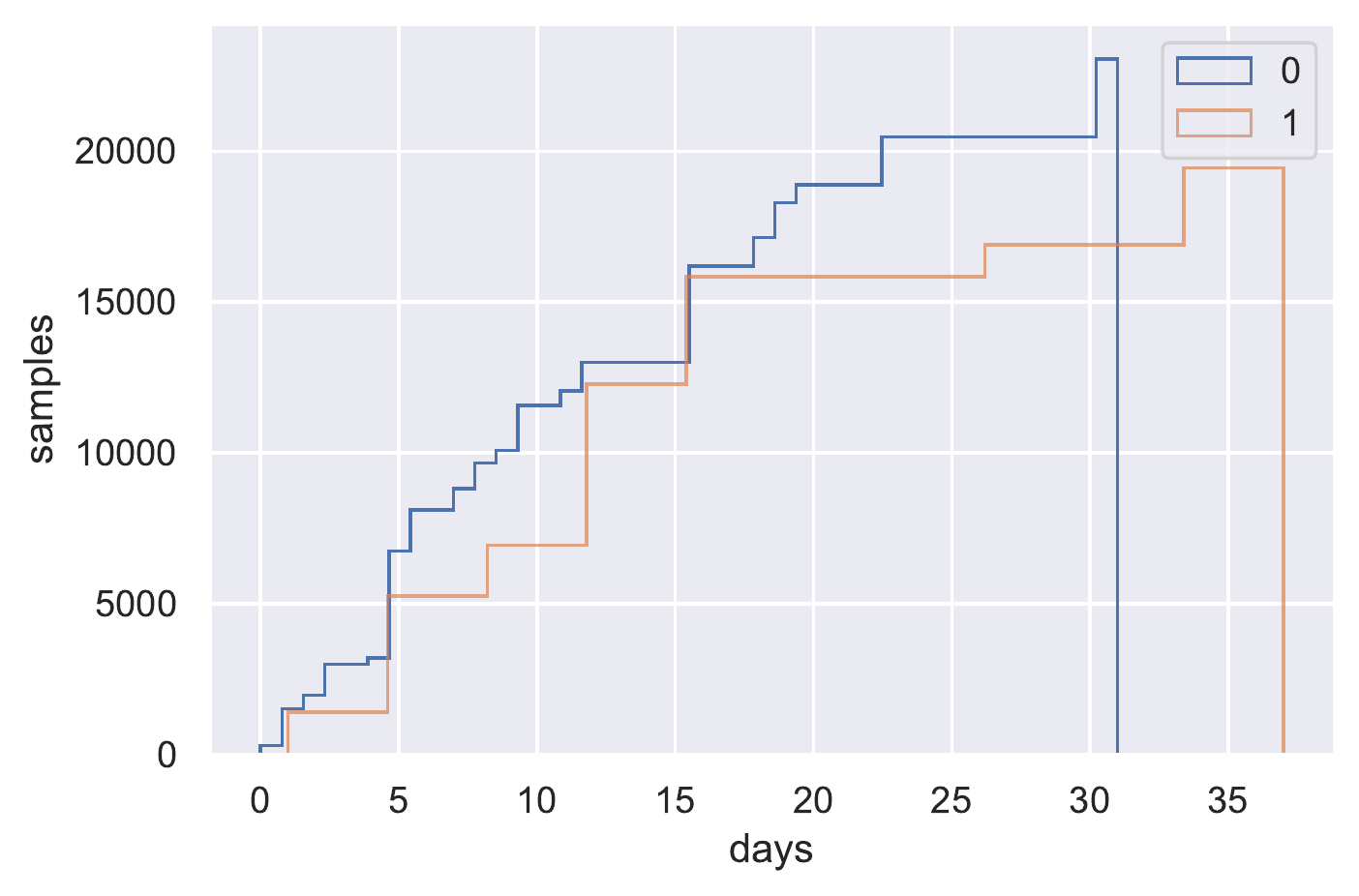}
\caption{Cumulative number of samples per day for each test group.}
\label{fig:cummu_samples}
\end{figure}

\begin{table}[!h]
\centering
\caption{Age distribution of subject with different COVID-19 test results.}
\label{tbl:age}
\resizebox{0.9\columnwidth}{!}{%
\begin{tabular}{|c|c|c|c|c|}
\hline
            & \multicolumn{2}{c|}{Number of Participants} & \multicolumn{2}{c|}{Number of Samples} \\
            \hline
 Age range  & Negative & Positive & Negative & Positive \\
\hline
21 - 40& 5 & 3 & 2650 & 1913\\
41 - 60& 10 & 11 & 6556 & 6528\\
61 - 80& 20 & 14 & 12430 & 7518\\
80 - 100& 3 & 4 & 1421 & 3490\\
\hline
\end{tabular}
}
\end{table}
In addition, another valuable aspect of this data set is the longitudinal monitoring of vital signs. Fig. \ref{fig:duration} shows the distribution of available data (i.e., blood pressure and heart rate) duration in days per each test result group. Throughout the remainder of this paper, the value of 0 represents negative, and 1 shows positive results.
Besides, Fig. \ref{fig:cummu_samples} shows the cumulative number of samples per day for each test group. 

\subsection{Ethics}
The data was jointly reviewed by the Institutional Review Boards of all UC Health campuses and was determined to be non-human subjects research. Moreover, UC-CORDS does not contain any patient identifier such as name and phone number. However, all original service dates (e.g., the date of the COVID-19 test) are preserved, and partial address information is available (i.e., town or city, state, and zip code). As such, UC-CORDS is a HIPAA Limited Data Set.

\subsection{Neural Networks}
\begin{figure*}[!th]
\centering
\includegraphics[width=\textwidth]{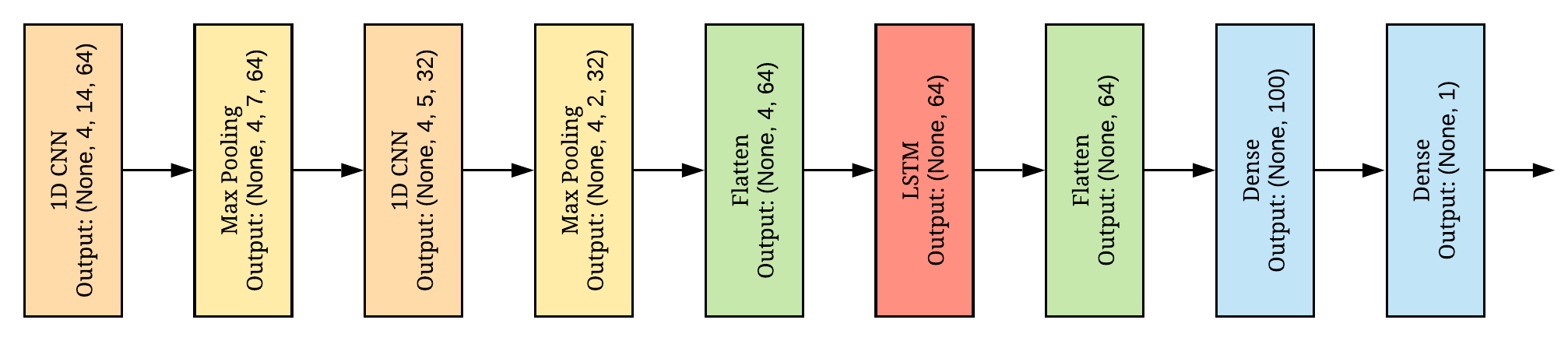}
\caption{The architecture of the proposed neural network.}
\label{fig:arch}
\end{figure*}

In this study, we were interested in the COVID-19 test result prediction using longitudinal heart rate and blood pressure monitoring. To perform the prediction, we used a deep neural network architecture combining Convolutional and Recurrent Neural Networks (CNN and RNN). Such a model was utilized to leverage the embedded structure of longitudinal data. We have considered three channels of vital signs, i.e., heart rate (HR), systolic and diastolic blood pressure (SBP and DBP), as the inputs and the test result for the network's output. Fig. \ref{fig:arch} illustrates the detailed structure of the proposed network. It consists of two 1-dimensional CNN, following by a max-pooling layer, a long short-term memory (LSTM) \cite{hochreiter1997long} layer, and finally two fully connected dense layers. We randomly selected 80\% of patients as train and the rest as the test data.

We labeled positive COVID-19 test results with 1 (13918 samples in the train, 43.86\%, and 3483 samples in the test data, 55.28\%) and the negative ones with 0. To perform the learning and testing tasks, TensorFlow package of Python has been utilized.

Besides, to see the prediction's effectiveness, we tested our model on different time intervals on test subjects. In other words, we were interested to see the possibility of test result prediction by only looking at a couple of samples (in days). Different test samples with different lengths (starting from 2 days of data until 28 days) have been extracted to answer this question.

Finally, for visualization purposes, the t-SNE method \cite{maaten2008visualizing} has been used over the output of the dense layer of the neural network to reduce the feature space's dimension to two.

\subsection{Statistical Analyses}
To show the correlation of features (i.e., blood pressures and heart rates) and the test results, statistical features have been extracted. We measured basic features, including mean, minimum (min), maximum (max), and standard deviation (std) of DBP, SBP, and HR. Besides, we utilized the Point Biserial correlation between the proposed features and the test results. This correlation is similar to Pearson’s correlation and is used when one of the variables is binary, and the other variable is a continuous number \cite{sheskin2020handbook}.

\section{Data Availability}
This study's data set is not publicly available as they contain protected patient health information and are institutional property.

\section{Code Availability}
The code developed for this study may be made available upon request for non-commercial use.

\section{Acknowledgment}
The project described was supported by the National Center for Research Resources and the National Center for Advancing Translational Sciences, National Institutes of Health, through Grant (UL1 TR001414). 
Charles A. Downs is also supported by NR016957. The content is solely the responsibility of the authors and does not necessarily represent the official views of the NIH.

\section{Author Contributions}
Study design: M.A.M., A.M.R., C.A.D; Data analysis: M.A.M.; Data interpretation: M.A.M., S.A.H.A, I.A., C.A.D., A.M.R; manuscript preparation: all authors; literature review: S.A.H.A, M.A.M; supervision: N.D., A.M.A; all authors reviewed the results and commented on the paper.

\section{Competing Interests}
The authors declare no competing interests.

\bibliographystyle{IEEEtran}
\bibliography{scibib}

\begin{thebibliography}{10}
\providecommand{\url}[1]{#1}
\csname url@samestyle\endcsname
\providecommand{\newblock}{\relax}
\providecommand{\bibinfo}[2]{#2}
\providecommand{\BIBentrySTDinterwordspacing}{\spaceskip=0pt\relax}
\providecommand{\BIBentryALTinterwordstretchfactor}{4}
\providecommand{\BIBentryALTinterwordspacing}{\spaceskip=\fontdimen2\font plus
\BIBentryALTinterwordstretchfactor\fontdimen3\font minus
  \fontdimen4\font\relax}
\providecommand{\BIBforeignlanguage}[2]{{%
\expandafter\ifx\csname l@#1\endcsname\relax
\typeout{** WARNING: IEEEtran.bst: No hyphenation pattern has been}%
\typeout{** loaded for the language `#1'. Using the pattern for}%
\typeout{** the default language instead.}%
\else
\language=\csname l@#1\endcsname
\fi
#2}}
\providecommand{\BIBdecl}{\relax}
\BIBdecl

\bibitem{li2020acute}
X.~Li and X.~Ma, ``Acute respiratory failure in covid-19: is it “typical”
  ards?'' \emph{Critical Care}, vol.~24, pp. 1--5, 2020.

\bibitem{tang2020comparison}
X.~Tang, R.~Du, R.~Wang \emph{et~al.}, ``Comparison of hospitalized patients
  with ards caused by covid-19 and h1n1 [published online ahead of print, 2020
  mar 26],'' \emph{Chest}, pp. 30\,558--4, 2020.

\bibitem{jhcovid}
``{COVID-19 dashboard by the Center for Systems Science and Engineering (CSSE)
  at Johns Hopkins University},''
  \url{https://gisanddata.maps.arcgis.com/apps/opsdashboard/index.html\#/bda7594740fd40299423467b48e9ecf6},
  [Online; accessed 31-Aug-2020].

\bibitem{jehi2020individualizing}
L.~Jehi, X.~Ji, A.~Milinovich, S.~Erzurum, B.~Rubin, S.~Gordon, J.~Young, and
  M.~W. Kattan, ``Individualizing risk prediction for positive covid-19
  testing: results from 11,672 patients.'' \emph{Chest}, 2020.

\bibitem{callahan2020estimating}
A.~Callahan, E.~Steinberg, J.~A. Fries, S.~Gombar, B.~Patel, C.~K. Corbin, and
  N.~H. Shah, ``Estimating the efficacy of symptom-based screening for
  covid-19,'' \emph{NPJ digital medicine}, vol.~3, no.~1, pp. 1--3, 2020.

\bibitem{force2012acute}
A.~D.~T. Force, V.~Ranieri, G.~Rubenfeld, B.~Thompson, N.~Ferguson, E.~Caldwell
  \emph{et~al.}, ``Acute respiratory distress syndrome,'' \emph{Jama}, vol.
  307, no.~23, pp. 2526--2533, 2012.

\bibitem{shashikumar2017early}
S.~P. Shashikumar, M.~D. Stanley, I.~Sadiq, Q.~Li, A.~Holder, G.~D. Clifford,
  and S.~Nemati, ``Early sepsis detection in critical care patients using
  multiscale blood pressure and heart rate dynamics,'' \emph{Journal of
  electrocardiology}, vol.~50, no.~6, pp. 739--743, 2017.

\bibitem{matsumura2020comparison}
K.~Matsumura, Y.~Toyoda, S.~Matsumoto, Y.~Kawai, T.~Mori, K.~Omasa, T.~Fukada,
  M.~Yamada, T.~Kazamaki, S.~Furugori \emph{et~al.}, ``Comparison of the
  clinical course of covid-19 pneumonia and acute respiratory distress syndrome
  in 2 passengers from the cruise ship diamond princess in february 2020,''
  \emph{The American journal of case reports}, vol.~21, pp. e926\,835--1, 2020.

\bibitem{gattinoni2020covid}
L.~Gattinoni, D.~Chiumello, and S.~Rossi, ``Covid-19 pneumonia: Ards or not?''
  2020.

\bibitem{UCCORD}
``{University of California Health creates centralized data set to accelerate
  COVID-19 research},''
  \url{https://www.universityofcalifornia.edu/press-room/university-california-health-creates-centralized-data-set-
  \ accelerate-covid-19-research}, 2020, [Online; accessed 15-Aug-2020].

\bibitem{omron}
``{Wearable Blood Pressure Monitor and Watch, HeartGuide by OMRON},''
  \url{https://omronhealthcare.com/}, [Online; accessed 31-Aug-2020].

\bibitem{maaten2008visualizing}
L.~v.~d. Maaten and G.~Hinton, ``Visualizing data using t-sne,'' \emph{Journal
  of machine learning research}, vol.~9, no. Nov, pp. 2579--2605, 2008.

\bibitem{wu2020risk}
C.~Wu, X.~Chen, Y.~Cai, X.~Zhou, S.~Xu, H.~Huang, L.~Zhang, X.~Zhou, C.~Du,
  Y.~Zhang \emph{et~al.}, ``Risk factors associated with acute respiratory
  distress syndrome and death in patients with coronavirus disease 2019
  pneumonia in wuhan, china,'' \emph{JAMA internal medicine}, 2020.

\bibitem{pasteur}
I.~Pasteur, ``{Whole genome of novel coronavirus, 2019-nCoV, sequenced},''
  \url{www.sciencedaily.com/releases/2020/01/200131114748.htm}, 31 January
  2020, [Online; accessed 31-Aug-2020].

\bibitem{bhatraju2020covid}
P.~K. Bhatraju, B.~J. Ghassemieh, M.~Nichols, R.~Kim, K.~R. Jerome, A.~K.
  Nalla, A.~L. Greninger, S.~Pipavath, M.~M. Wurfel, L.~Evans \emph{et~al.},
  ``Covid-19 in critically ill patients in the seattle region—case series,''
  \emph{New England Journal of Medicine}, vol. 382, no.~21, pp. 2012--2022,
  2020.

\bibitem{goyal2020clinical}
P.~Goyal, J.~J. Choi, L.~C. Pinheiro, E.~J. Schenck, R.~Chen, A.~Jabri, M.~J.
  Satlin, T.~R. Campion~Jr, M.~Nahid, J.~B. Ringel \emph{et~al.}, ``Clinical
  characteristics of covid-19 in new york city,'' \emph{New England Journal of
  Medicine}, 2020.

\bibitem{richardson2020and}
S.~Richardson, J.~Hirsch, M.~Narasimhan, J.~Crawford, T.~McGinn, and
  K.~Davidson, ``and the northwell covid-19 research consortium presenting
  characteristics, comorbidities, and outcomes among 5700 patients hospitalized
  with covid-19 in the new york city area,'' \emph{JAMA}, vol.~10, 2020.

\bibitem{carlino2020predictors}
M.~V. Carlino, N.~Valenti, F.~Cesaro, A.~Costanzo, G.~Cristiano, M.~Guarino,
  and A.~Sforza, ``Predictors of intensive care unit admission in patients with
  coronavirus disease 2019 (covid-19),'' \emph{Monaldi Archives for Chest
  Disease}, vol.~90, no.~3, 2020.

\bibitem{grasselli2020baseline}
G.~Grasselli, A.~Zangrillo, A.~Zanella, M.~Antonelli, L.~Cabrini, A.~Castelli,
  D.~Cereda, A.~Coluccello, G.~Foti, R.~Fumagalli \emph{et~al.}, ``Baseline
  characteristics and outcomes of 1591 patients infected with sars-cov-2
  admitted to icus of the lombardy region, italy,'' \emph{Jama}, vol. 323,
  no.~16, pp. 1574--1581, 2020.

\bibitem{wang2020clinical}
D.~Wang, B.~Hu, C.~Hu, F.~Zhu, X.~Liu, J.~Zhang, B.~Wang, H.~Xiang, Z.~Cheng,
  Y.~Xiong \emph{et~al.}, ``Clinical characteristics of 138 hospitalized
  patients with 2019 novel coronavirus--infected pneumonia in wuhan, china,''
  \emph{Jama}, vol. 323, no.~11, pp. 1061--1069, 2020.

\bibitem{SNOMED}
``{SNOMED Clinical Terms},'' \url{https://www.snomed.org/}, [Online; accessed
  15-Aug-2020].

\bibitem{hochreiter1997long}
S.~Hochreiter and J.~Schmidhuber, ``Long short-term memory,'' \emph{Neural
  computation}, vol.~9, no.~8, pp. 1735--1780, 1997.

\bibitem{sheskin2020handbook}
D.~J. Sheskin, \emph{Handbook of parametric and nonparametric statistical
  procedures}.\hskip 1em plus 0.5em minus 0.4em\relax crc Press, 2020.

\end{thebibliography}

\end{document}